\input epsf
\documentstyle[twocolumn,aps,epsf]{revtex}
\begin{document} 
\twocolumn[\hsize\textwidth\columnwidth\hsize\csname 
@twocolumnfalse\endcsname
\title{Structure and Vibrations of the Vicinal Copper (211) Surface}
\author{C.Y. Wei$^1$, Steven P. Lewis$^2$, E.J. Mele$^1$ and Andrew M. Rappe$^2$}
\address{ $^1$Department of Physics \\$^2$Department of Chemistry \\
Laboratory for Research on the Structure of Matter
\\University of Pennsylvania
\\Philadelphia, PA  19104 
}
\maketitle
\begin{abstract} We report a  first-principles theoretical study of the surface relaxation and lattice dynamics 
of the Cu (211) surface using the plane-wave pseuodopotential method. 
 We find large atomic relaxations for the first several atomic layers near the step edges
 on this surface, and a substantial step-induced renormalization of the surface harmonic force
 constants. We use the results to study the harmonic fluctuations around the equilibrium structure and identify three new step-derived features in the zone-center vibrational
 spectrum. Comparisons of these results with previous theoretical work and with experimental
 studies using inelastic He atom  scattering are reported.  
\end{abstract}  
 ]

\section{Introduction}

	At a vicinal surface the average surface normal is tipped away from a high
symmetry low-index direction, and the microscopic structure consists of an ordered array
of atomic steps which separate low-index terraces.  
The microscopic atomic structure near a step edge is very different from that found
 in the bulk or at a low-index surface, and as a consequence one expects a substantial redistribution
 of the valence electronic charge density near the step edges. This can play an important
 role in modifying the local dynamical properties and the chemical reactivity.  Therefore  stepped surfaces
are important systems to study exprimentally and theoretically. In fact
 the influence  of steps on the surface chemistry can be so large that the
 macroscopic reactivity of many real surfaces is governed by dynamical phenomena at
 steps. Understanding these processes at the microscopic level can lead to control of
 surface dynamical phenomena through control and design of the surface structure. 

	Due to the importance of these structures,
the dynamics of stepped surfaces have been the subject of several experimental investigations using modern surface-science techniques within the last several years. Witte and coworkers
 have studied the effects of surface steps on the surface vibrational spectra of Cu(211)
 and Cu(511) using inelastic He atom scattering \cite{witte}. These studies were able to resolve clearly the
 zone-folding effects in the low frequency surface vibrational spectra due to the ordering of single atom
 steps on these surfaces.  However, this work did not  address the renormalization
 of the interatomic interactions which occur near the step edges, and the authors concluded that their
 observations could be interpreted within a model in which the parameters describing the
 surface atomic interactions were assigned their bulk values. More recently Niu et al.
 \cite{Niu} studied the dynamics of Ni(977) using inelastic He atom scattering. This work
 revealed a new step-derived branch in the surface vibrational spectrum which the 
 authors associated with a new kind of surface sound wave, bound to the surface along the
 surface normal, and localized laterally to the surface step-edge. To understand the dynamics
 of this step-induced branch, these authors introduced a phenomenological model in 
 which the step edge modes were interpreted as the modes of a stretched string running
 parallel to the surface.  Using this model, the authors found that the effective elastic
 constants describing the harmonic fluctuations of the string were strongly renormalized from the
 values appropriate for bulk Ni or for a low-index Ni surface. However, subsequent
theoretical work, based on a continuum model
 for a semi-infinite elastic medium, demonstrated that the step-derived features can be understood
 using continuum dynamics and enforcing the surface boundary condition
 appropriate to a  weakly corrugated stepped surface \cite{Mele}. Thus the changes of the surface
 interatomic interactions near the step edges remains an important but unresolved
 issue. 

	In this paper we report a study of the surface structure and dynamics of the Cu(211)
surface using first-principles pseudopotential theory. Previously, the equilibrium 
structures for various  stepped vicinal surfaces have been examined theoretically using empirical
 interaction potentials \cite{witte}, tight binding theory \cite{Papa}, effective
 medium theory \cite{Rahman} and first-principles density functional theory \cite{Nelson}. The former three 
methods are only tested in a highly coordinated bulk
 environment; the transferability of these approaches to the structures
 found on  vicinal surfaces is thus open to question. First-principles total-energy calculations
 provide a parameter free method for studying equilibrium structures. Unfortunately the
 structural complexity of vicinal surfaces has made them nearly inaccessible  by first-principles 
 methods, particularly for the transition metals. Nonetheless several years ago Nelson and Feibelman \cite{Nelson}
  reported the first \it{ab initio} \rm study of the equilibrium structure of a simple metallic vicinal surface,
 Al(331). These authors
 found that the atomic relaxations near the step edges are significantly enhanced with respect
 to those found at close-packed low-index surfaces, although the dynamical properties of the  Al(331) surface
 were not investigated in their work.  Recent improvements in computational algorithms and 
hardware now allow the detailed exploration of realistic models for vicinal
surfaces of transition metals using first-principles total-energy methods.  

	In this paper we apply the first-principles pseudopotential theory to study the
equilibrium structure and dynamics of the Cu(211) surface.    We find that  atomic
relaxations for Cu(211)  are strongly enhanced in the first several layers near the
 surface step edges. This relaxation strongly changes the effective surface harmonic
 force constants from the values appropriate for the bulk or for a low-index surface. 
  Nevertheless, we find that the dominant  effect on the surface vibrational spectra is simply the 
 lowering of the surface translational symmetry and the folding of the surface Brillouin 
 zone at the stepped surface. Specifically, in our calculations we find three primary step-induced features in the vibrational spectrum at the surface Brillouin
 zone center.  Two of these  are the ``backfolded" versions of the elementary surface
 elastic waves of the close packed (111) surface, while 
 one is due to  the renormalization of the surface
 interatomic force constants.  One of our calculated low-energy modes  is  in
 good agreement with   an excitation observed by inelastic He scattering experiments on Cu(211) \cite{witte},  whereas
 the symmetry and energies of the other two features made them  inaccessible in  these measurements.
 In this paper we suggest other experimental probes which could be used to examine these latter
 excitations.

	The remainder of this paper is organized as follows. In Section II we summarize the
 symmetry and structural features  of the Cu(211) surface, and in  Section III we briefly present
 our computational methodology. The equilibrium structure predicted in our calculations is presented
 in Section IV. In Section V we apply the model to determine the harmonic interlayer force
 constants near the surface, and  use them to examine the surface vibrational excitations. In Section VI we
 analyze  the step-induced features in the surface vibrational spectra. A  discussion and summary
 of the results is presented in section VII.  

\section{Structure of the Copper (211) Surface}

	The ideal lattice termination of the (211) surface of an fcc 
crystal is shown in  the top panel of Figure 1. A more complete geometric classification of the 
 atomic structure for stepped surfaces of fcc crystals  with
 close-packed step terraces is given in Reference \cite{black}. At the (211) surface
 the surface normal  is rotated by 19.5$^\circ$ away from  the (111) direction with
 the axis of rotation along the $ (01 \bar 1)$ direction   exposing
 (111) terraces containing 3 atoms  separated by single atom (100) facets at the step edges.
 The Cu(211) surface has a primitive rectangular unit cell corresponding to a 3
 $\times$ 1 reconstruction of the bulk lattice structure.   
 The surface retains a mirror plane symmetry perpendicular to the 
  $ (01 \bar 1)$ direction. Although this surface cell exhibits a microscopic faceted
 structure at the atomic scale, the unit cell is relatively compact and can be studied theoretically using conventional thin film
 supercell methods. The calculations reported below are carried out on a structural
 model in which a thin film containing 17 atoms in the primitive cell is terminated
 on its (211) and $(\bar 2 \bar 1 \bar 1 )$ planes. An illustration of this structural model, with the inequivalent
 lattice sites identified, is shown in  the bottom panel of Figure 1. 

\section{Computational Methods}

	The calculations are carried out using  density functional theory within the local
 density approximation (non-relativistic)\cite{JDJ}. The electron-ion potential for Cu is modeled using an optimized non-local
 soft  core pseudopotential for the Cu $11+$ ion. Cu, like other first row 
 transition metals, is challening for  pseudopotential theory because
 it possesses an unscreened 3$d$ valence shell.  This implies that the peak in the 3$d$-derived
 valence charge density is relatively tightly bound to the core, and thus  a well converged
 calculation  requires a large  energy cutoff in the plane wave basis set. This 
 difficulty has been addressed and solved by the introduction of optimized pseudopotentials,
 which allow a well converged calculation of the 3d derived levels with only  a modest energy
 cutoff in the plane wave basis set. A discussion  of the construction and testing of these
 potentials can be found in Reference \cite{AMR}. The reference state for our potential
 is the $+1$
 ionic configuration  for Cu with orbital occupations $4s^{0.75}4p^{0.25}3d^9$. The one-electron wave
 functions are expanded in a plane-wave basis set, up to a kinetic energy cutoff of 50 Ry. 
 Using this cutoff  we find that the truncation error in the computed total energy is less than 1 mRy per atom.  
 In the 17 layer thin film shown in Figure 1, the thin film structure is repeated periodically along  the $x$ 
 direction with a 14.5 $\rm \AA$ vacuum layer separating the neighboring films. For this structure approximately 25,000
 plane waves are required in the basis set for the electronic degrees of freedom. We minimize the energy with respect to the
 plane-wave expansion coefficients using a pre-conditioned conjugate gradient algorithm
 described in detail in Reference \cite{JDJ}. 

	Using this method for the bulk fcc structure in copper we calculate
 an equilibrium nearest-neighbor separation of 2.554 $ \rm \AA$, which agrees quite well
 with the experimental value of 2.553 $\rm \AA$.  As a test
 of the lattice dynamical properties of Cu using this potential, we also carried out  frozen-phonon calculations of the (100) zone boundary transverse and
 longitudinal acoustic phonons in the  bulk;
 the calculated (experimental) energies  are: TA: 32.2 (30.2) meV,
 and LA: 22.8 (20.7) meV \cite{exphonon}. The bulk modulus is $1.63 \times 10^{11}$ 
 $\rm N/m^2$ which is larger than, although  comparable to, the observed value $1.42 \times 10^{11}$
 $\rm N/m^2$ \cite{Geschneider}. All these results are in good agreement with experiment
 and in excellent agreement with several previous theoretical studies of the bulk elastic properties of Cu within
 the local density approximation \cite{rodach,troullier,chelik,zunger}. Our
calculations are nonrelativistic, that is we have not included scalar relativistic corrections in our theory. These are known to introduce a slight (1 $\%$ - 1.5 $\%$) additional contraction in the predicted lattice constant, and a 
 corresponding increase in the bulk modulus.

	To study the harmonic fluctuations around the equilibrium structure, we compute
 the distribution of forces  at  all of the  lattice sites of the thin film produced in 
 response  to  separate  small 
 displacements of each of the atoms away from their equilibrium locations.
 The maximum ionic displacement that we studied in these calculations is 0.05 $\rm \AA$. 
Using the Hellmann-Feynman theorem, 
 the induced force on the  $i$-th site in the $\alpha$-th polarization 
 is $F_{i, \alpha} = -  {\partial U} / {\partial x_{i,
\alpha}} = - \langle {\partial H} / {\partial x_{i, \alpha}
 \rangle}$ where  $H$ is the Kohn-Sham Hamiltonian for the coupled electron-ion system, and the brackets denote an expectation value in the relaxed electronic ground state
 in the presence of the ionic displacement.
 The harmonic interaction constant coupling the
$i \alpha$ and $j \beta$  displacements is $K_{ \alpha,  \beta}(i,j) = {\partial^2 U} / {\partial x_{i \alpha} \partial x_{j \beta}} = - {\partial F_{i \alpha}} / {\partial x_{j \beta}}$ .  
 The displacement fields and forces calculated by this method  describe the dynamics
 at the center of the folded Brillouin zone for the stepped surface, and so in the
 analysis below we consider only the vibrational spectra at $\bar{\Gamma}$.

 	The harmonic interlayer force constants calculated  from our thin film  describe the physics 
 only within the region near the surface (half the thickness of our film). This is
 insufficient to describe realistically the dynamics of a semi-infinite  crystal
  terminated on the (211) plane. To remedy this problem, we carry out a separate 
 calculation to obtain the analogous interlayer force constants for  Cu in  its bulk fcc structure. We then
 construct a composite model in which the surface region is recoupled to a very thick  unperturbed 
 bulk film containing $\approx$ 1000 atomic layers. This expanded film is sufficiently large to allow us to 
 locate   critical points in the bulk phonon spectra accurately and to represent  the bulk and
 surface modes of our system properly. Since the dynamical matrix is computed at the center of the
 surface Brillouin zone,
  mirror reflection through the  $(0 1 \bar 1)$ plane remains a valid symmetry, and
  therefore the dynamical matrix can be  decoupled into two sectors describing lattice modes which are even and
 odd under reflection through this mirror plane.

\section{Equilibrium Structure}

	Figure 2 illustrates the atomic structure of the (211) surface of our fcc crystal. 
 The arrows represent magnified displacements of each of the atoms
 from its position in the ideal termination of the bulk Cu lattice to its position in
 the equilbrium structure (for clarity these displacements have been magnified by roughly a factor of 20).
 By symmetry the atomic displacements in the relaxed structure are confined to the
plane perpendicular to the step edge (i.e. the  $( 0 1 \bar 1 ) $ plane).
   We find that the lateral displacements, i.e. along the $(\bar 111)$ direction, while
 allowed by symmetry, are all relatively small; the largest  of these corresponds to $\approx 1.5 \% $ of the
 lateral spacing in the bulk. By contrast the displacements along the surface normal
 are quite substantial near the surface. The relaxation of the distance 
between the  two outermost atomic layers is $-14.4 \% $
 of the bulk interlayer spacing in this direction.   For the  fully relaxed geometry
 the fractional change of the nearest-neighbor distance from the step edge atoms
 is nearly $ 4 \% $ of the bulk nearest-neighbor spacing. For comparison, at the
 Cu(111) surface the inward relaxation of the top surface layer reduces the nearest-neighbor
 spacing by $ 0.8 \% $ and for Cu(100) the nearest-neighbor spacing is reduced by $ 1 \%$.
 Thus for the vicinal surface the atomic relaxations along the surface normal are substantially
  larger than obtained for a close packed low index surface. The relaxation within the 
 surface plane  is quite small. 

The displacement
 pattern shown in Figure 2 illustrates the Smoluchowski effect \cite{Sm}  which  
  predicts a reduction of the height of the step edge in the equilibrium structure. Qualitatively
 the conduction electron density leaks away from the crystal at the step edge,
 tending  to smooth the step edge profile. It is easy to see that 
 this generates an electric field which  pulls the ion at the inside corner of the step
 edge away from the thin film  
 and pushes the ion at the outside corner towards the film. A detailed 
 tabulation of the calculated atomic displacements is given in Table I. In addition these  results show that
 the structure obtained near the center of the slab is essentially the bulk fcc structure,
 and so the film thickness is sufficient to describe the relaxation of the isolated 
 (211) surface accurately.

\section {Interlayer Force Constants}

	To study the harmonic fluctuations around the equilibrium relaxed structure, we calculate
the interlayer force constants in our film following the method outlined in Section III.  
The atomic relaxation at the surface leads to very large modifications of the harmonic
 interlayer force constants near the surface region.  In Table
 II we report  a set of selected interatomic interactions near the surface, and compare these with
 their  corresponding unperturbed bulk values. We can clearly identify the following 
 trends. The interlayer coupling constants for displacements polarized \it along the surface
 normal  \rm is much larger at the stepped surface than in the bulk. For example the 
 harmonic interaction constant  which couples  the displacement of surface site 1 to the nearest-neighbor subsurface site, $K_{xx}(1,4)$, is 4.52 $\rm  eV/\AA^2$ at the surface while the comparable interaction is  2.89 $\rm eV/\AA^2$
 in the bulk. The intra-terrace interactions are also strongly renormalized at
 the surface; as an example  we find that $K_{xx}(1,2)$ = 0.41 $\rm eV/\AA^2$ at the surface terrace, which is to be compared with 
 the  bulk value  0.12 $\rm eV/\AA^2$.  For displacements polarized within the surface 
 plane, the surface force constants are again larger, although the changes of the force constants are  more modest, and quite comparable to those obtained  for a flat low-index surface \cite{steve}. For example  
  the surface (bulk) interaction constant  $K_{zz}(1,2)$ = 3.24(2.70) $\rm eV/\AA^2$ and
 $K_{zz}(1,4)$ = -0.49(-0.21) $\rm eV/\AA^2$.  Note that the latter represents a very large fractional renormalization
 of the surface interaction; this is mainly due to the fact that the \it interlayer \rm
 coupling between displacements parallel to the surface plane is relatively small in 
 both environments.  The renormalization of the surface interaction constants for displacements
 parallel to the step direction is similar,e.g. $K_{yy}(1,2)$ = 1.27(0.92) $\rm eV/\AA^2$
 and $K_{yy}(1,4)$ = 1.47(0.92) $\rm eV/\AA^2$.  Generally we find that all the interaction
 constants are strongly enhanced near the step edges at the surface and this effect  tends to offset  the
 rather large  decrease in the coordination number for the surface species. This is most
 clearly seen by inspecting the diagonal interaction constants for atoms  on 
a common terrace. For example,
  the  interactions $K_{xx}(i,i)$  for the three sites $i$ 
 at the surface terrace have the values -5.77, -4.61 and
 -4.37 $\rm eV/\AA^2$ in the relaxed structure for sites $i$=1,2,3. Site 1 is the site at the  outside step corner
 with the smallest coordination, and site 3 is the site at the inside corner with the
 largest coordination  for a terrace site.

\section{Surface Vibrational Spectra}

	We combine the interlayer force constants for the Cu(211) surface region with the 
 values calculated for bulk Cu to obtain a composite dynamical matrix for a thick Cu film which
 terminates on  a (211) lattice plane. The dynamical matrix is decoupled into two sectors 
 describing the modes of the combined systems which are even and odd, respectively,  under
 reflection through the $(0  \bar 1 1)$ mirror plane.  For a slab containing $N$ atoms, the dimensions of the even and odd subspaces are $2N$ and $N$ respectively. 

	In Figures 3 and 4  we present results for the  symmetry-projected densities of vibrational 
states as a function of frequency at the  $\bar{\Gamma}$-point in the
surface Brillouin zone.  Figure 3 gives the density of modes which are even
 under reflection through the surface mirror plane.
 The thin solid curve is obtained from a sum over all the modes
 of the composite slab, while the bold solid curve gives the \it local \rm density of states $
 	n_s(\omega) = \sum_n \sum_{i, \alpha} | \phi_{i \alpha,n} |^2 \delta (\omega - \omega_n)$
 where  $\phi_{i \alpha, n}$ is the amplitude of the $n$-th normal mode 
 on site $i$ with polarization $\alpha$, and 
 where the sum over $ i \alpha $ runs  over the symmetry allowed displacements of the
 three atoms on the outermost surface terrace, and the sum over $n$ runs  over all the modes of the
 slab.  

The surface projected density of even states clearly reveals two step-induced features in the
 surface vibrational spectrum at $\bar{\Gamma}$. The lower of these, at an energy of 11.5 meV,
 is a backfolded image of the Rayleigh wave of the ``flat" (211) surface
  which would appear at a momentum $q$ = 1.005 $\rm \AA^{-1}$ along the $(\bar 1 1 1)$ direction
 in a continuum theory. 
  We find that  the $q$ = $\pm$   1.005 $\rm \AA^{-1}$ backfolded 
branches are strongly
 mixed by the cell periodic variations in the 
 surface dynamical matrix so that only the lower branch remains localized below
 the nearby  bulk transverse  threshold at 12.4 meV. This excitation  is  a surface ``standing wave" trapped
 on the surface terrace. This can be seen more clearly in the map of the displacement pattern
 for this feature plotted in Figure 5.  It shows that  the largest atomic displacements occur  at the center 
 of the surface terraces and  are reduced at the step edges, an effect which can be seen
  clearly in the subsurface displacement field.  Nevertheless this  mode possesses a large amplitude at the surface, polarized along the surface normal, and so this branch  is expected to be  strongly allowed  in
 inelastic atom scattering from the surface.  Indeed, the calculated frequency agrees closely with the 10.9 meV
 feature observed in inelastic He scattering experiments from Cu(211) \cite{witte}.

	The backfolded Rayleigh branch overlaps the projected bulk continuum at  $\bar{\Gamma}$, yet the
 mode remains quite sharp in the surface-projected vibrational spectrum. Within a continuum elastic theory, mixing between the backfolded
 Rayleigh mode and the projected bulk modes is dynamically forbidden at $\bar{\Gamma}$, since in the continuum theory  the coupling
 between the two is mediated by the surface strain which vanishes for the lowest branch
 of bulk acoustic waves at $\bar {\Gamma}$ \cite{Mele}.  This dynamical symmetry  can be  weakly
 violated in a lattice model. However, the sharpness of this surface-derived feature in the
 vibrational spectrum shows that the mixing remains very weak, even for the relatively narrow
 terraces found on Cu (211).  

	At higher energy we find a second strong step induced surface feature
  at 24.7 meV which corresponds to a backfolded image of the surface longitudinal resonance of a flat surface.   The
 displacement field associated with this mode is illustrated
in Figure 6. Note  that the surface displacements have their largest amplitudes polarized mainly parallel to the plane containing the surface terraces. This mode can also  
 be understood as a surface standing wave which is trapped laterally by repeated scattering from the step 
 edges. However, it is apparent from the plot of the displacement field, that the effect of
 scattering from the step edges is much weaker than one finds for the lower energy backfolded
 Rayleigh mode. This is because this higher energy feature is not a true
 surface-localized mode, but is resonantly mixed with  propagating bulk acoustic waves which penetrate
 infinitely deeply into the bulk. Since the displacement field for this mode is polarized mainly within the surface plane  we expect it to be a poor candidate for study by inelastic atom scattering,
 which is most sensitive to atomic displacements along the surface normal. Nevertheless, it is
 very interesting that, unlike the situation for a flat low index surface, a
 linear coupling between the backfolded longitudinal resonance and charge fluctuations along
 the surface normal is allowed by symmetry in this system since both excitations are even with respect to 
 the $(0 1 \bar 1)$ mirror plane.  This opens the possibility of probing this excitation by
 inelastic scattering of a charged probe (e.g. high resolution inelastic electron scattering)
 from the surface.  Further theoretical work is needed to quantify the magnitude of the charge
 fluctuations coupled to this mode. We expect that the important charge fluctuations
 coupled to the zone folded longitudinal resonance will be found at the step edges, and that
 this coupling can be used to probe the dynamics of the conduction electrons near the steps.  

	Figure 4 shows the zone center vibrational spectra for modes that are odd with respect to the $(0 1 \bar 1)$ mirror plane. 	The thin solid curve represents the spectrum summed over all the modes of the
 film, while the bold curve gives the modes weighted by the squares of the displacement fields
 in the surface terrace.  Here  the step induced features, though present, are much
 more subtle. A broad transverse resonance, labeled TA2, is found centered
 at  11.3 meV just below the bulk 
 projected critical point.  The displacement field for this excitation is plotted 
 in Figure 7. We see that the mode corresponds to a transverse shear wave which  is 
 strongly laterally confined at the step edges. This  mode is completely absent within a simple continuum elastic theory, 
 and its appearance in the spectrum directly reflects the rather substantial renormalization of the interatomic force constants
 within the surface region.  This mode is a poor candidate for study using neutral-atom
 scattering since, as shown in Figure 7, it is  polarized 
 in the surface plane, along the $(0 1 \bar 1)$ direction and is odd under the surface mirror plane. Likewise
 linear coupling between this mode and z-polarized charge fluctuations is symmetry forbidden,
 and therefore this mode will be silent in charged-particle scattering as well.  

 \section {Discussion}

	In this section we compare our results to experimental measurements of the surface
 phonon spectra on the Cu(211) surface and to related previous theoretical work. 

 	Witte et al. \cite{witte} used inelastic He scattering to study the 
 phonons of the Cu(211) surface.  These authors experimentally studied the surface
vibrational excitations below 15 meV as a function of momentum transfer
 in the surface plane, both parallel and perpendicular to the surface steps.
 At the surface Brillouin zone center, $\bar \Gamma$, an excitation polarized
 along the surface normal at an energy of 10.9 meV was observed and assigned to the 
 backfolded image of the Rayleigh wave of a flat surface. This agrees quite well with the
 Rayleigh mode we calculate at 11.5 meV. The experiments do not extend
 to sufficiently high energy to detect the backfolded LA resonance, which
we find at 24.7 meV. Furthermore, as we commented in the previous section, 
 the  11.3 meV TA2 resonance has atomic displacements \it parallel \rm to the
 surface plane, and would normally be undetectable by neutral-atom scattering.

 	 These authors also report a  study of the surface
 dynamics using a central nearest-neighbor harmonic force constant model. 
 Qualitatively, this simple model correctly reproduces the zone-center surface
 excitations which we have calculated. The agreement between the two
 theories is at first surprising in view of the rather large changes
 of  the surface harmonic interactions which we  presented in Tables 2 and 3. However, in the long wavelength limit these surface features are ``leaky" 
 excitations which penetrate well below the first atomic layer of the surface.
 The calculated frequency of such an excitation is relatively insensitive to the
 precise value of the surface interatomic interactions, and instead depends
 on the fact that propagating elastic waves will be perfectly reflected 
 from the surface. Close to the surface Brillouin zone boundaries the 
 effective penetration depth of these modes is substantially smaller, and the
 frequencies are much more sensitive to the surface interatomic potentials.  

	More recently Durukanoglu, Kara and Rahman  (DKR)\cite{Rahman} reported a theoretical study
 of the equilibrium structure and dynamics of Cu vicinals using the embedded atom
 method  (EAM) \cite{foiles}. For Cu(211), these authors also observed enhanced structural relaxations
 near the step edges as we do, and the magnitudes of their predicted relaxations are
  consistent with the trends
 we display in Table I. They also compute interatomic force constants near
 the step edges, and find that the \it interatomic  \rm force constants are enhanced near the
 step edges. DKR also note that the \it diagonal \rm elements of the dynamical
 matrix are decreased relative to their bulk values; this follows directly from 
 the lower coordination of the surface atoms.  Thus the off-diagonal terms provide
 a useful quantitative measure of the enhancement of the ``back bonding"  of under-coordinated atoms
 near the step edges.  Although the EAM and LDA results both indicate an enhancement
 of these interactions, the enhancement is approximately twice as large  within the LDA than
 that obtained with the EAM model. The local density approximation provides a more complete
 description of the relaxation of the conduction charge density, particularly near
 the step edges. In our view this quantitative discrepancy between the calculated force constants
 reflects the fact that these depend rather sensitively on the relaxation of the
 charge density near the step sites. 

	Our calculations point to an important piece of new physics at the stepped surface.
 As we emphasized in Section VI, the surface lattice dynamics is dominated by 
 a simple zone folding effect by which the finite momentum Rayleigh wave, and longitudinal
 resonance are folded back 
 through the surface Brillouin zone center $\bar {\Gamma}$.   Normally these excitations
 are observed as modes with finite momentum transfer in the surface plane and
 are accessible only by neutral atom scattering or by inelastic electron scattering
 in the impact regime. On the vicinal surface, these backfolded modes can be  coupled to
 charge fluctuations near the surface, an effect which we expect to be particularly strong near the  step edges.
  Thus these excitations can be coupled  to a macroscopic charge fluctuation of the surface
   dipole layer, and are accessible by inelastic electron scattering.
 This is particularly significant because inelastic electron scattering can  be used to study the 
 dynamics of the conduction charge near step edges, and to examine the effects of adsorbates,
 step wandering, etc. on this coupling. These are central issues in the surface chemistry
 associated with the step edges, and we are therefore currently investigating this effect theoretically
 by calculating the strength of this coupling.  

	In summary, we have reported the first \it{ab-initio} \rm study of the structure and
dynamics of a stepped surface for a transition metal. We find that atomic relaxations near
the step edges for the Cu(211) surface are substantially enhanced with respect to their counterparts at a low
 index surface. This produces  very large changes of the surface harmonic force
 constants, much larger than  inferred from previous experiment and theory. 
 For the Cu(211) surface we find that there are three important step-derived excitations
 in the Brillouin zone center vibrational spectrum.  Our result for the frequency of one mode,
corresponding to a  
backfolded Rayleigh wave agrees well with experimental measurements using inelastic He
 scattering.  The two other features were inaccessible to these measurements, and 
 we have suggested a method for observing these other excitations.

	This work was supported by the NSF under grants DMR 93 13047 and
DMR 97 02514.

\begin{table}
\caption{Summary of data for the calculated equilibrium structure of
 Cu(211). The data give the  percentage changes in interatomic spacings 
 with respect to the ideal termination of the bulk 
 fcc lattice along a (211) plane. $d_{ij}$ label interlayer distances
 along the surface normal, and $l_{ij}$ label distances parallel
 to the surface plane, along the $(\bar 1 1 1)$ direction. }

\begin{tabular}{cdcd}
 \multicolumn{2}{c}{Relaxation ($\%$)}&\multicolumn{2}{c}{Registry ($\%$)} \\
\tableline
$d_{12}$&-14.4 &$l_{12}$&-1.6 \\
$d_{23}$&-10.7 &$l_{23}$&-1.0  \\
$d_{34}$&10.9 &$l_{34}$&-0.7  \\
$d_{45}$&-3.8 &$l_{45}$& 1.4  \\
$d_{56}$&-2.3 &$l_{56}$&-0.4  \\
$d_{67}$& 1.7 &$l_{67}$& 0.3   \\
$d_{78}$&-1.0 &$l_{78}$& 0.2 \\
$d_{89}$&-0.05 &$l_{89}$&-0.03 \\
\end{tabular}
\end{table}

\begin{table}
\caption{Selected interlayer surface harmonic interaction constants are 
 compared with their values in the bulk. The atoms  and the
 polarization of the displacements are labeled
 according to the notation of Figure 1. All constants are in units of 
 $\rm eV/\AA^2$.}
\begin{tabular}{cccc}
 &Polarization&$K_{\alpha \alpha}$(1,4)&K$_{\alpha \alpha}$(1,2) \\
\tableline
surface&$\alpha$ = $x$&4.52 & 0.41\\
bulk&$\alpha$ =  $x$& 2.89& 0.12\\
\\
surface&$\alpha$ =  $y$& 1.47& 1.27\\
bulk&$ \alpha$ = $y$ &0.92& 0.92\\
 \\
surface&$\alpha$ = $z$& -0.49& 3.24\\
bulk&$\alpha$ = $z$& -0.21& 2.70\\
\end{tabular}
\end{table}
 \begin{table}
\caption{Diagonal components of the harmonic force constants for
 the surface terrace atoms on Cu(211). The values of $K_{xx}(i,i)$ are given
 in $\rm eV/\AA^2$ for the three atoms along the surface terrace, using
 the labeling convention of Figure 1.}
\begin{tabular}{ccc}
$K_{xx}(1,1) $&$K_{xx}(2,2)$&$K_{xx}(3,3)$ \\
\tableline
-5.77&-4.61&-4.37 \\
\end{tabular}
\end{table}

\begin{figure*}[hbt]
\epsfxsize=3in
\centerline{\epsfbox[31 31 583 761]{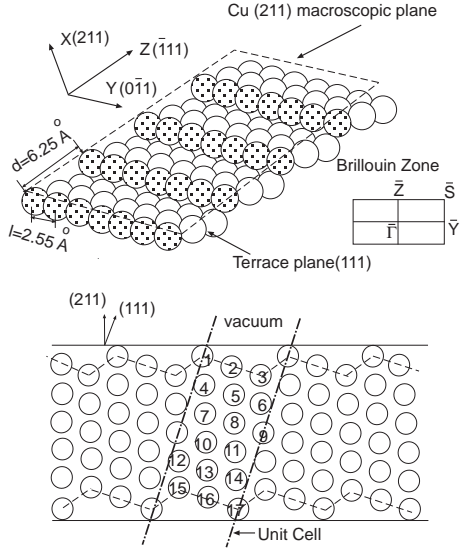}}
\caption{Structure of the Cu(211) vicinal surface. The surface consists
 of three-atom (111) terraces separated by single atom (100) steps. The 
 lattice parameters and conventions for  our coordinate system are
 illustrated in the top panel. We model the (211) surface with a film
 containing 17 Cu sites. The lower panel is a projection of this film on the 
 $\rm (0 1 \bar 1)$ plane with the atoms in the unit cell identified.}
\end{figure*}
\newpage
 
\begin{figure*}[hbt]
\epsfxsize=3in
\centerline{\epsfbox[31 31 583 761]{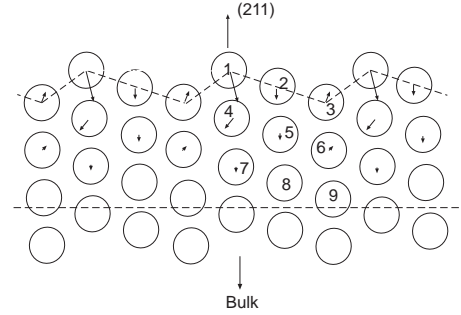}}
\caption{Atomic displacements in the equilibrium structure of our
 model thin film for Cu(211).  The arrows show the \it magnified \rm displacements
 of each atom from  its position in the ideal lattice termination to the equilibrium position. For clarity the displacements have been magnified by roughly
 a factor of 20.}
\end{figure*}

\begin{figure}
\epsfxsize=3in
\centerline{\epsfbox[31 31 583 761]{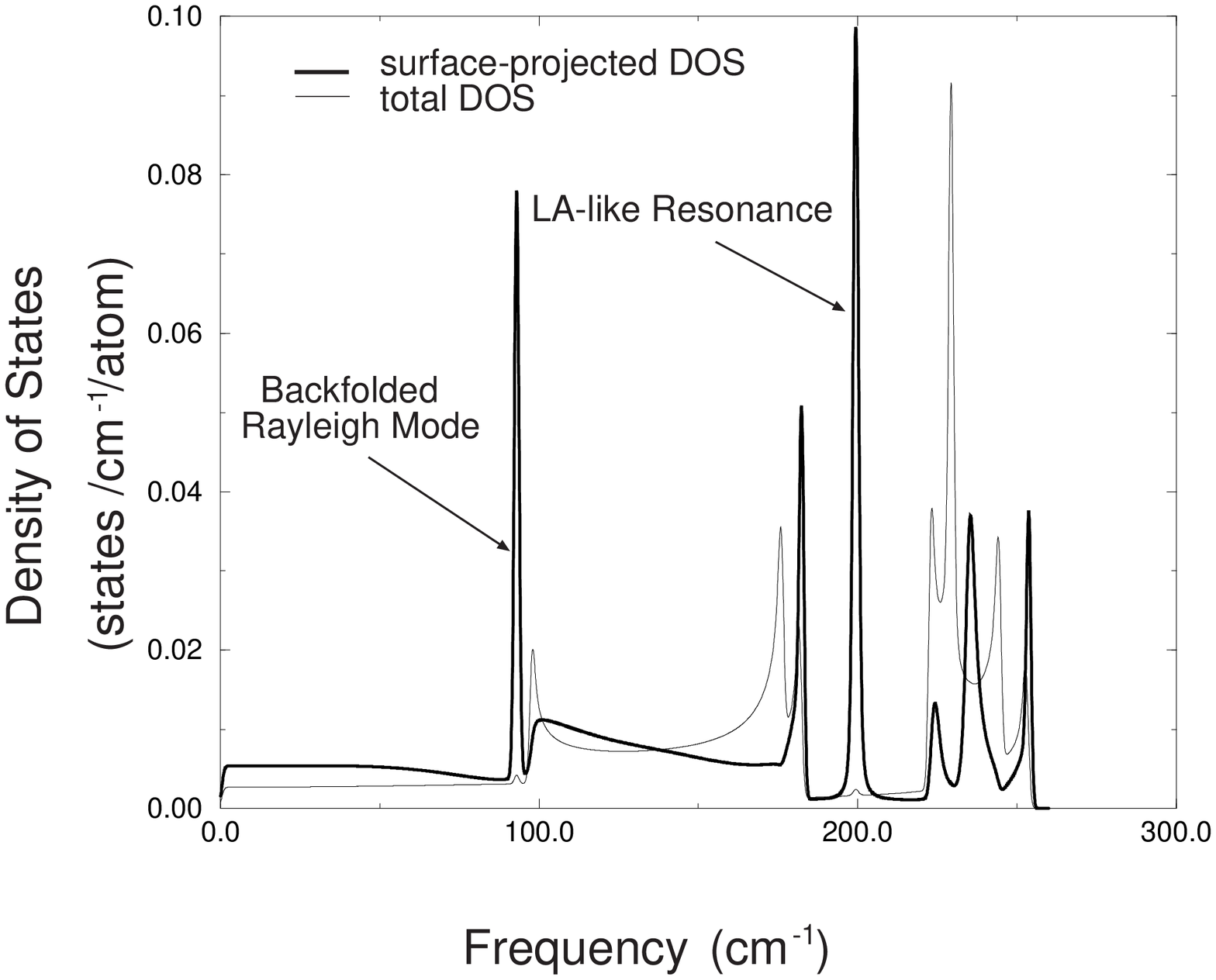}}
\caption{Vibrational densities of states for the Cu(211) film. The spectra
 give the densities of states for modes which are even under
 reflection through the $(0  \bar 1 1)$ mirror plane. The light curve is a trace
 over  even modes of the expanded film. The bold curve is a trace 
 over the  modes weighted by the squares of the normal-mode amplitudes
 in the surface terrace.} 
\end{figure}
 
\begin{figure}
\epsfxsize=3in
\centerline{\epsfbox[31 31 583 761]{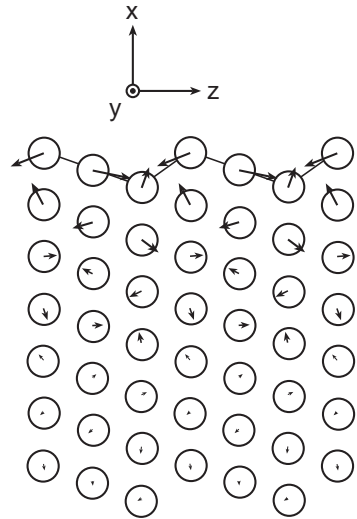}}
\caption{Vibrational densities of states for the Cu(211) film. The spectra
 give the densities of states for modes which are odd under
 reflection through the $(0  1 \bar 1)$ mirror plane. The light curve is a trace
 over  odd  modes of the expanded film. The bold curve is a trace 
 over the  modes weighted by the squares of the normal-mode amplitudes
 in the surface terrace.}
\end{figure}

\begin{figure}
\epsfxsize=3in
\centerline{\epsfbox[31 31 583 761]{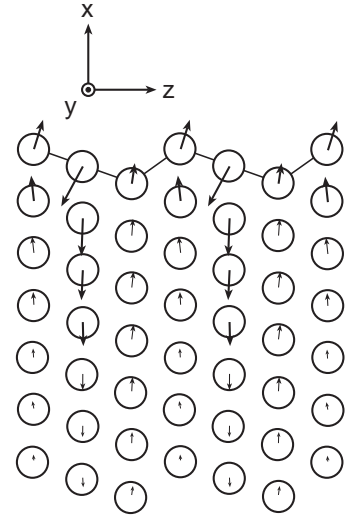}}
\caption{A snapshot of the displacement field for the backfolded 
Rayleigh mode at 11.5 meV in the surface vibrational spectrum. The
amplitudes decay exponentially into the bulk. At the surface, this
 mode has a standing wave  pattern which is ``pinched" at the step edges.}
\end{figure}

\begin{figure}
\epsfxsize=3in
\centerline{\epsfbox[31 31 583 761]{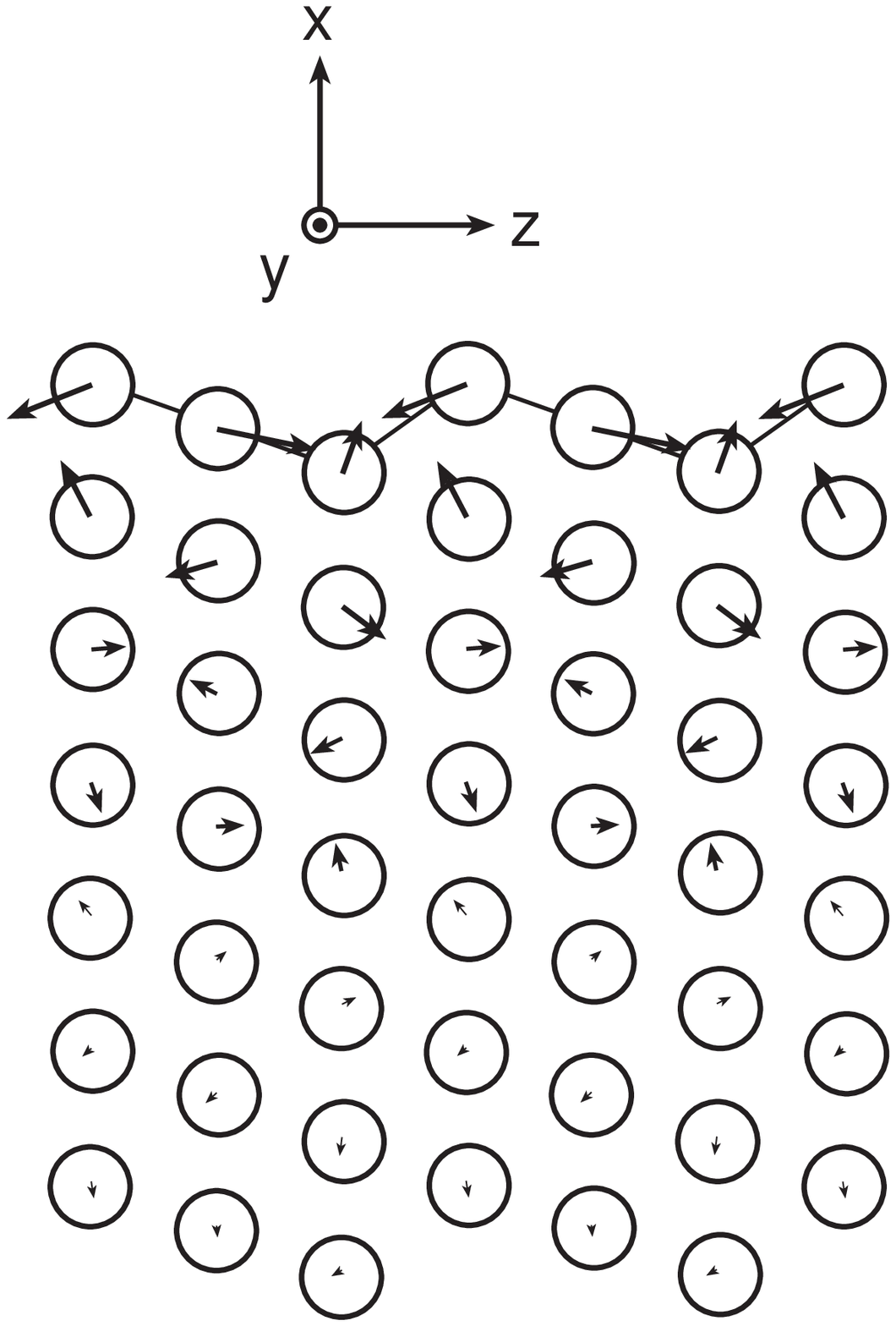}}
\caption{ A snapshot of the displacement field for the  backfolded longitudinal resonance centered at 24.7 meV. The mode is weakly resonant with bulk excitations. At the surface, the
 atomic displacements are mainly polarized parallel to the surface.}
\end{figure}

\begin{figure}
\epsfxsize=3in
\centerline{\epsfbox[31 31 583 761]{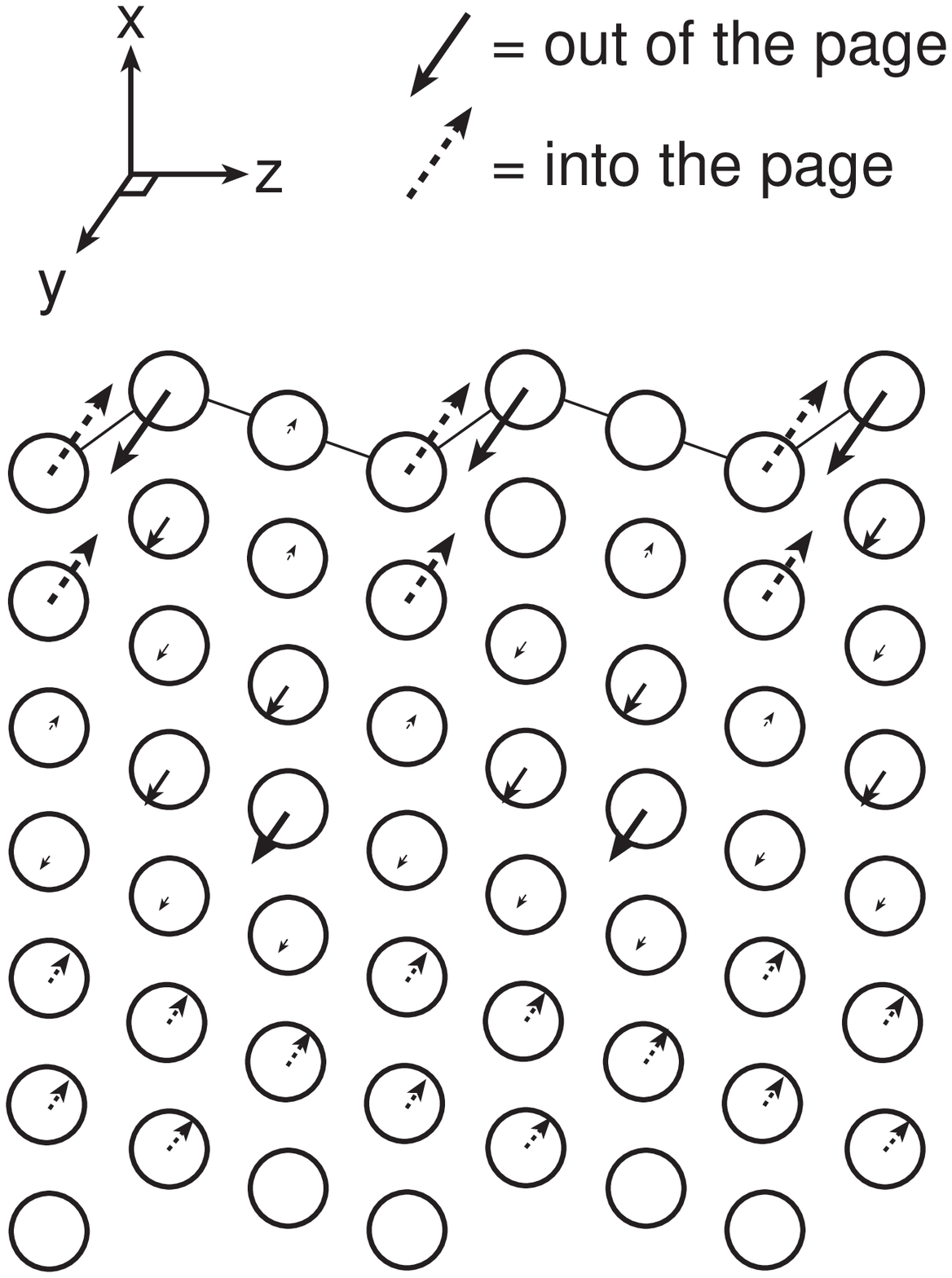}}
\caption{A snapshot of the displacement field for the TA2
resonance centered at 11.3 meV. This mode is odd under reflection through the  $(0  1 \bar 1)$
 mirror plane, and can be described as a shear horizontal mode polarized
 parallel to the surface steps and localized at the surface step edge.}
\end{figure}


\begin{references}
\bibitem{witte} G. Witte, J. Braun, A Lock and J.P. Toennies,
Physical Review B {\bf 52}, 2165 (1995)
\bibitem{Niu} L. Niu, D. Koleske, D. Gaspar and S. Sibener, Journal
 of Chemical Physics
{\bf 102}, 9077 (1995)
\bibitem{Mele} E.J. Mele and M.V. Pykhtin, Physical Review Letters {\bf 75},
3878 (1995)
\bibitem{Papa} S. Papadia, M.C. Desjonqueres, D. Spanjaard, Physical
 Review B {\bf 53}, 4083 (1996)
\bibitem{Rahman} S. Durukanoglu, A. Kara and T.S. Rahman, Physical Review B {\bf 55},
13 894 (1997)
\bibitem{Nelson} J.S. Nelson and P.J. Feibelman, Physical Review Letters,
{\bf 68}, 2188 (1992)
\bibitem{black} J.E. Black and P. Bopp, Surface Science {\bf 140}, 275 (1984)
\bibitem{JDJ} M.C. Payne, M.P. Teter, D.C. Allan and J.D. Joannopoulos,
Reviews of Modern Physics {\bf 64}, 1045 (1992)
\bibitem{AMR} A.M. Rappe, K.M. Rabe, E. Kaxiras and J.D. Joannopoulos,
Physical Review B {\bf 41}, 1227 (1990)
\bibitem{exphonon} R.M. Nicklow, G.Gilat, H.G. Smith, L.J. Raubenheimer,
 and M.K. Wilkinson, Physical Review {\bf 164}, (1967)
\bibitem{Geschneider} K. Geshneider, Jr. in \it Solid State Physics, \rm
H. Ehrenreich, F. Seitz and D. Turnbull, eds., (Academic, New York, 1964), Vol. 16
\bibitem{rodach} Th. Rodach, K-P Bohnen, and K-M Ho, Surface Science {\bf 286}, 66 (1993)
\bibitem{troullier} N. Troullier and J.L. Martins, Physical Review B {\bf 43}, 1993 (1991)
\bibitem{chelik} J.R. Chelikowsky and M.Y. Chou, Physical Review B {\bf 38}, 7966 (1988)
\bibitem{zunger} Z.W. Lu, S-H Wei and A. Zunger, Physical Review B {\bf 41}, 2699 (1990)
\bibitem{Sm} R. Smoluchowski, Physical Review {\bf 60}, 661 (1941)
\bibitem{steve} One finds that interlayer force constants at the relaxed
 Cu(001) surface are renormalized by $\lesssim 20 \%$ with respect to their
 bulk values. S.P. Lewis and A.M. Rappe (unpublished).
\bibitem{foiles} S.M. Foiles, M.I. Baskes and M.S. Daw, Physical Review B {\bf 33}, 7983 (1986)
\end{references}
\end{document}